\documentclass[amstex]{article}
\usepackage{graphicx}
\usepackage{dcolumn}
\usepackage{amsmath}
\usepackage{amssymb}
\usepackage{amsfonts}
\usepackage{amscd}

\newtheorem{theorem}{Theorem}
\newtheorem{lmm}[theorem]{Lemma}

\newtheorem{pro}[theorem]{Proposition}
\newtheorem{df}[theorem]{Definition}
\newtheorem{rmk}[theorem]{Remark}

\newcommand\calA{{\cal A}}

\newcommand\calB{{\cal B}}
\newcommand\calC{{\cal C}}
\newcommand\calD{{\cal D}}

\newcommand\calK{{\cal K}}

\newcommand\calM{{\cal M}}
\newcommand\calN{{\cal N}}

\newcommand\bfZ{\bf Z}

\newcommand\dprime{\prime\prime}

\newcommand{\abs}[1]{\left|#1\right|}

\begin{document}
\newpage\thispagestyle{empty}
\begin{center}
{\Huge\bf
On Haag Duality for 
\\
Pure States of Quantum Spin Chain}
\\
\bigskip\bigskip
\bigskip\bigskip
{\Large M. Keyl}
\\
Institute for Scientific Interchange Foundation,
\\
Viale S. Severo 65 - 10133 Torino, Italy
\\
m.keyl@tu-bs.de
\\
\bigskip\bigskip
{\Large Taku Matsui}
\\
 Graduate School of Mathematics, Kyushu University,
\\
1-10-6 Hakozaki, Fukuoka 812-8581, JAPAN
\\
 matsui@math.kyushu-u.ac.jp
\\
\bigskip\bigskip
{\Large   D. Schlingemann* and R.~F. Werner**}
\\
Institut f{\"u}r Mathematische Physik, TU Braunschweig,
\\
Mendelssohnstr.3, 38106 Braunschweig, Germany.
\\
*d.schlingemann@tu-bs.de, 
\\
** r.werner@tu-bs.de
\\
\bigskip\bigskip
March, 2007
\end{center}
\bigskip\bigskip\bigskip\bigskip
\bigskip\bigskip\bigskip\bigskip
{\bf Abstract:}  In this note, we consider quantum spin chains and their translationally invariant pure
states. 
We prove  Haag duality  for quasilocal observables  localized 
in semi-infinite intervals $(\infty , -1]$ and $[0,\infty )$
when the von Neumann algebra generated by observables localized in $[0,\infty )$ is non type I.
\\
\\
{\bf Keywords:} UHF algebra, pure state, translational invariance, Haag duality, Cuntz algebra.
\\
{\bf AMS subject classification:} 82B10 

\newpage
\section{Introduction.}\label{Intro}
\setcounter{theorem}{0}
\setcounter{equation}{0}
In local Quantum Field Theory, the Haag duality
is a crucial notion in structure analysis.
(See \cite{Haag}. ) In this note, we consider the Haag duality for
quantum spin systems on a one-dimensional lattice in an irreducible
representation. By Haag duality we mean that the von Neumann algebra
${\frak M}_{\Lambda}$  generated by observables localized in 
an infinite subset  $\Lambda$ of $\bfZ$ is the commutant of the von Neumann algebra  
${\frak M}_{\Lambda^{c}}$ generated by observables localized in the complement
$\Lambda^{c}$ of $\Lambda$. This duality plays a crucial role in analysis of entanglement
property of states of infinite spin chain. See \cite{KSW} and \cite{KMSW}.
\par
If these von Neumann algebra ${\frak M}_{\Lambda}$ is of type I, 
the duality is very easy to show.
However, even if the representation of a whole quasi-local algebra
is irreducible, the restriction to an infinite region may give rise to 
a non-type I von Neumann sub-algebra. For example, the restriction of the ground state of 
massless XY model to  the semi-infinite interval $[0, \infty)$ gives rise to a type III 
von Neumann algebra and  we believe that the same is true for 
the spin $1/2$ massless antiferromagnetic XXZ chain.
Though Haag duality  is a basic concept, it seems that the proof of Haag duality
is not obtained so far for the general case when both $\Lambda$ and its complement 
$\Lambda^{c}$ are infinite sets. We will see that the duality holds when
the representation contains a vector state which is translationally invariant
and $\Lambda=[1,\infty)$.
\\
\\
To explain our results more precisely, we introduce our notation now. 
By $\frak A$, we denote the UHF $C^*-$algebra $d^{\infty}$
(the infinite tensor product of d by d matrix algebras ) :
$${\frak A} = \overline{\bigotimes_{\bfZ} \: M_{d}({\bf C})}^{C^*} . $$
Each component of the tensor product above is specified with a lattice
site $j \in \bfZ$.
By $Q^{(j)}$ we denote the element of ${\frak A}$ with $Q$ in
the jth component of the tensor product and the identity in any other
component.
For a subset $\Lambda$ of $\bfZ$ , ${\frak A}_{\Lambda}$ is defined as
the $C^*$-subalgebra of ${\frak A}$ generated by elements $Q^{(j)}$ 
with all $j$ in $\Lambda$.
We set
$${\frak A}_{loc} = \cup_{ \Lambda \subset {\bfZ} : | \Lambda | < \infty}
 \:\: {\frak A}_{\Lambda}  $$
where the cardinality of $\Lambda$ is denoted by $|\Lambda |$.
We call an element of ${\frak A}_{loc} $ a local observable
or a strictly local observable.
\par
When $\varphi$ is a state of ${\frak A}$, the restriction of $\varphi$
to ${\frak A}_{\Lambda}$ will be denoted by $\varphi_{\Lambda}$ :
$$\varphi_{\Lambda} = \varphi \vert_{{\frak A}_{\Lambda}} .$$
We set 
$${\frak A}_R =  {{\frak A}}_{[1,\infty)} \: , \:
{\frak A}_L =  {\frak A}_{(-\infty, 0]} \: , \:  \varphi_R=\varphi_{[1,\infty)} \: , \:
\varphi_L = \varphi_{(-\infty, 0]}  \: \: .$$
By $\tau_j$ we denote the automorphism
of $\frak A$ determined by $ \tau_j(Q^{(k)})=Q^{(j+k)}$ for any j and k in $\bfZ$.
$\tau_j$ is referred to as the lattice translation of ${\frak A}$.
\par
Given a representation $\pi$ of $\frak A$ on a Hilbert space,
the von Neumann algebra generated by $\pi({\frak A}_{\Lambda})$
is denoted by ${\frak M}_{\Lambda}$. We set
$${\frak M}_R = {\frak M}_{[1,\infty)} =\pi ({\frak A}_{R})^{\dprime} ,  \quad
{\frak M}_L = {\frak M}_{(-\infty , 0]} =\pi ({\frak A}_{L})^{\dprime} .$$
For a state $\psi$ of a $C^{*}$-algebra $\cal A$ we denote the GNS triple by
$\{ \pi_{\psi} ({\frak A}), \Omega_{\psi} , {\frak H}_{\psi} \}$ where $ \pi_{\psi}$ is 
the GNS representation and $\Omega_{\psi} $ is the GNS cyclic vector 
in  the GNS Hilbert space $\pi_{\psi}$. 
\begin{theorem}
Let $\varphi$ be a translationally invariant pure state of the UHF algebra
$\frak A$. and let $\{\pi_{\varphi} (\frak A), \Omega_{\varphi} , {\frak H}_{\varphi}  \}$ 
be the GNS triple for  $\varphi$.
Then, the Haag duality holds:
\begin{equation}
{\frak M}_R = {\frak M}_L^{\prime}
\label{eqn:a1}
\end{equation}
\label{th:duality}
\end{theorem}
\begin{rmk}
\label{rmk:type}
 We consider the situation that the state may not be faithful.
In  Proposition 4.2 \cite{KMSW}, we have shown that ${\frak M}_R$ appearing in our context
cannot be a type $II_1$ factor.
Precise statement and its proof is included here in Lemma \ref{lmm:type1factor} .
We are not aware of any example of  ${\frak M}_R$
which is of type $II_{\infty}$.  
${\frak M}_R$ is of type $III$ in generic cases. For example,
when the state $\varphi_{R}$ is faithful, ${\frak M}_R$ is of type $III_{1}$ due to Theorem 4
of \cite{Longo2} by R.Longo.
More precisely,
 An endomorphism $\hat{\tau}$ of a factor is strongly asymptotically
abelian if
$$ \lim_n [ \hat{\tau}^n (Q) , R ] =0$$
in strong operator topology for any $Q$ and $R$ in $\frak M$.
 If $\varphi_R$ is faithful, the restriction of the (normal extension)
shift $\tau_1$ of ${\frak M}$ to ${\frak M}_R$ is an strongly asymptotic abelian endomorphism 
of ${\frak M}_R$
and  ${\frak M}_R$ is a type $III_{1}$ factor. See \cite{Longo2}.
\end{rmk}
\begin{rmk}
\label{rmk:BJKR}
In our proof of Haag duality, we consider a gauge invariant extension
of the state $\varphi$ to a state of the tensor product $O_d \otimes O_{d}$ of Cuntz algebras  
and show the Haag duality at this level.
 We use ideas of \cite{BJKW} in our proof ,though,
 our way of proof is different from  \cite{BJKW}. In  \cite{BJKW},
O.Bratteli, P.Jorgensen, A.Kishimoto and R.Werner
focus on dilation of Popescu systems to representations of Cuntz algebras and their 
pure states while our starting point is a pure state of 
(two-sided infinite) UHF algebras and go down to Popescu systems.  
\\
 At first look, the section 7 of the paper  \cite{BJKW}  may give an impression
  that Proof of Theorem 7.1 of  \cite{BJKW} implies Haag duality.
 (c.f. Lemma 7.7 and Lemma 7.8)
 However,  for the KMS state of the standard $U(1)$gauge action of $O_{d}$ 
 the assumption of Theorem 7.1 of  \cite{BJKW} are satisfied 
 both  Lemma 7.7 and Lemma 7.8 do not hold. 
 \\
 Let $S_{j}$ be the Cuntz generator and consider the gauge action $\gamma_{z}$
 defined in Section 2. Then the $\beta = \ln d$ KMS state $\psi$ 
 is unique , in particular it is faithful and the gauge invariant extension of 
 the trace of ${\frak M}_R$.
 Then the assumption of Theorem 7.1 of  \cite{BJKW} is satisfied for
 the GNS representation $\{\pi_{\psi}, {\frak H}_{\psi}\}$ of 
 $O_{d}$ associated with $\psi$ if we set  $\calK ={\frak H}_{\psi}$ .
  $V_{j} =\pi_{\psi}(S_{j})$. Then,  $\tilde{V}_{j} = \frac{1}{d}J\pi_{\psi}(S_{j}^{*})J$
and 
$${\frak H}_{\psi} ={\frak H}_{0} , \quad E |_{\calK}=P= I_{\calK} .$$ 
Nevertheless, the state $\omega$ is not pure and the equivalence of conditions (i) and (iii)
of Theorem 7.1 of  \cite{BJKW}  is valid.
\par
We prove that Lemma 7.6  of  \cite{BJKW}  is valid when the state of $\frak A$ is pure, 
and for that purpose we introduce new ideas in Section \ref{SecCuntz}.
Our ideas are based on the observation that   the translation $\tau_1$ is an inner automorphism
of  $O_d \otimes O_{d}$.
We do not use Commuting Lifting Theorem of   \cite{BJKW} for our proof of Haag duality.
\end{rmk}
\section{Split Property}\label{SecSplit}
\setcounter{theorem}{0}
\setcounter{equation}{0}
 One key word in our analysis is {\em split property} or {\em split inclusion}.
\par
Let $\frak M_{1}$ and $\frak M_{2}$ be factors acting on a Hilbert space $\frak H$ satisfying
$\frak M_{1} \subset \frak M_{2}^{\prime}$. We say the inclusion is split
if and only if  there exists an intermediate type I factor $\calN$ such that 
$\frak M_{1} \subset \calN \subset \frak M_{2}^{\prime}$. 
\\
The split inclusion is introduced for analysis of local QFT and of von Neumann algebras in 1980's.
(c.f. \cite{DoplicherLongo} )
In \cite{Longo1} R.Longo used the notion for his solution to the factorial Stone-Weierstrass
 conjecture. 
\\
If mutually commuting factors $\frak M_{1}$ and $\frak M_{2}$ acting on a Hilbert space $\frak H$
have a common cyclic and separating vector, say $\Omega$,
 the inclusion $\frak M_{1} \subset \frak M_{2}^{\prime}$ is called {\em standard}.
The standard split inclusion is a weak notion of independence of two quantum
systems. 
Let  $\varphi$ be the vector state associated with the common cyclic and separating 
vector $\Omega$ for $\frak M_{1}$ and $\frak M_{2}$.
A standard inclusion  $\frak M_{1} \subset \frak M_{2}^{\prime}$
is split if and only if $\varphi$ is quasi-equivalent to a product state
$\psi_{1}\otimes \psi_{2}$ where $\psi_{1}$(resp.$\psi_{2}$) is a normal state of
$\frak M_{1}$ (resp. $\frak M_{2}$) (c.f.\cite{Longo1}) .
In our case, $\frak M_{1} = \pi ( \frak A_{\Lambda})^{\dprime} $,  
$\frak M_{2}=\pi (\frak  A_{\Lambda^{c}})^{\dprime} $.
We note that $\frak M_{1}$ and $\frak M_{2}$ may not have a common cyclic and separating
vector in the GNS Hilbert space associated with a translationally invariant pure state and
 our inclusion may not be standard.

\begin{df}
A state $\varphi$ of the UHF algebra $\frak A$ for a one-dimensional quantum spin system
has split property with respect to $\Lambda$ and $\Lambda^{c}$ if and only if
$\varphi$ is quasi-equivalent to the product state $\varphi_{\Lambda}\otimes 
 \varphi_{\Lambda^{c}}$.
\label{df:split} 
\end{df}
It is easy to see that  $\varphi$ has the split property if and only if $\varphi$ is quasi-equivalent 
to another product state $\psi_{1}\otimes \psi_{2}$.
When  $\varphi$ is pure,   $\varphi$ is  unitarily equivalent to a pure product state and
the von Neumann algebra $\pi_{\varphi} ({\frak A}_{\Lambda})^{\dprime}$ is of type I.
Moreover when  $\varphi$ is pure, $\pi_{\varphi} ({\frak A}_{\Lambda})^{\dprime}$ is of type I,
if and only if $\varphi$ has the split property.
Thus if the von Neumann algebra ${\frak M}_{\Lambda}$
generated $\pi_{\varphi} ({\frak A}_{\Lambda} )$ is of type I, the Haag duality is very easy to see.
\begin{lmm}
\label{lmm:type1factor}
Let $\varphi$ be a pure state of $\frak A$. If  the von Neumann algebra ${\frak M}_{\Lambda}$
generated $\pi_{\varphi} ({\frak A}_{\Lambda} )$ is of type I, then
${\frak M}_{\Lambda} = {\frak M}_{\Lambda^{c}}^{'} .$
\end{lmm}
{\it Proof.}
 As the pure state $\varphi$ of $\frak A$ is split with respect to $\Lambda$ and    
$\Lambda^{c}$ , $\varphi$ is unitarily equivalent to 
$\psi_{1}\otimes \psi_{2}$ where $\psi_{1}$ (resp. $\psi_{2}$) is a state of 
${\frak A}_{\Lambda}$ (resp. ${\frak A}_{\Lambda^{c}}$).
The GNS Hilbert space $\frak H_{\varphi}$ associated with $\varphi$ is unitarily equivalent to
 the tensor product $\frak H_{\psi_{1}} \otimes \frak H_{\psi_{2}}$ and
${\frak M}_{\Lambda}= B({\frak H}_{\psi_{1}}) \otimes 1_{{\frak H}_{\psi_{2}}}, \quad 
{\frak M}_{\Lambda^{c}}= 1_{{\frak H}_{\psi_{1}}} \otimes B({\frak H}_{\psi_{2}} )
=  {\frak M}_{\Lambda}^{\prime} $.
{\it End of Proof.}
\bigskip
\bigskip
\par
As a consequence, in our proof of Haag duality, we concentrate on pure states $\varphi$
which are not quasi-equivalent to $\varphi_{\Lambda} \otimes \varphi_{\Lambda^{c}}$.
Existence of   a  translationally invariant pure without the split property for $\Lambda =[1,\infty)$, 
is highly non-trivial. In \cite{Split}, we have shown that ground states of some spin 1/2 systems 
satisfy these requirement. 
\\
\par
When $\varphi$ is a translationally invariant factor state of ${\frak A}$,
$\varphi_R$ gives rise to a shift of the von Neumann algebra
${\frak M}_R$ in the following way.
As there exists a unitary $U$ implementing the shift $\tau_{1}$ specified with
$U\pi (Q) \Omega_{\varphi} = \pi(\tau_{1}(Q))\Omega_{\varphi} $
for $Q$ in ${\frak A}$.
$Ad (U)$ gives rise to an endomorphism on the factor ${\frak M}_{R}$
 generated by $\pi_{\varphi} ({\frak A}_{R})$. We denote this endomorphism of ${\frak M}_{R}$
 by $\hat{\tau}_{1}$:
 $UQU^{*} =   \hat{\tau}_{1}(Q)$ ($Q \in {\frak M}_{R}$).
By definition,
$$\cap_{n=0}^{\infty} \hat{\tau}_1^n (\frak M_R) = {\bf C}1. $$
\begin{lmm}
\label{lmm:type2factor}
Let $\varphi$ be a translationally invariant pure state and let ${\frak M}_{R}$ be
the von Neumann algebra generated by $\pi_{\varphi} ({\frak A}_{R})$.
 ${\frak M}_{R}$ cannot be of type $II_{1}$.
\end{lmm}
{\it Proof.}
\\
Suppose that ${\frak M}_{R}$ is of type $II_{1}$ and let $tr$ be its unique normal tracial state.
The shift endomorphism of ${\frak A}_{R}$ is a limit of cyclic permutations of $(1,2,\cdots ,n)$ of lattice site
which is implemented by unitary $U_n$, $ \tau_1 (Q) = \lim U_n Q U_n^*$.
It turns out that the trace is invariant under $\hat{\tau}_1$ because
$$tr(\hat{\tau} \pi_{\varphi}(Q)) =  tr( \pi_{\varphi}(\tau_1 (Q)) = 
\lim_{n\to \infty}  tr( \pi_{\varphi}(U_n Q U_n^*))= tr( \pi_{\varphi}(Q ))$$
Thus, as $\varphi$ is the unique normal shift invariant state,
$\varphi_R = tr$. Then, the two sided translationally invariant extension of $tr$ to ${\frak A}$
is a trace and this contradicts with our assumption that $\varphi$ is pure.
{\it End of Proof.}
\\
\\
If a translationally invariant pure state $\varphi$ has the split property, 
the endomorphism $\Theta_{R}$ of $\frak A_{R}$ defined 
as the restriction of $\tau_{1}$ to $\frak A_{R}$ 
 is weakly inner on the GNS subspace associated with  $\varphi$.
 More precisely, let  $\Theta_{R}$ be an endomorphism of $\frak A$ determined by
 $\Theta_{R}(Q) = \tau_{1}(Q)$ for $Q \in \frak A_{R}$
 and $\Theta_{R}(Q) = Q$ for $Q \in \frak A_{L}$. 
\par
If $\varphi$ is  a translationally invariant pure state of $\frak A$ with the split property,
there exist isometries  $S_{j}$ ( $j = 1,2,\cdots ,d $) acting on the GNS space
associated with $\varphi$ satisfying generating relations of the Cuntz algebra
(c.f. the next section)
$ S_{j}^{*} S_{i}  =\delta_{ij} 1$ , $ \sum_{k=1}^{d}  S_{k}S_{k}^{*} = 1$
and  
\begin{equation}
\sum_{j=1}^{d}  S_{j}  \pi_{\varphi} ( Q)  S_{j}^{*}   = \pi_{\varphi}(\Theta_{R}(Q)), \quad
 S_{j} \in  \pi_{\varphi} (\frak A )^{\dprime}   \quad\quad \mbox{( $Q \in {\frak A}$).}
\label{eqn:e1}
\end{equation}  
As a consequence of weakly inner property of $\Theta_{R}$, 
$\varphi$ and $\varphi \circ \Theta_{R}$ are mutually quasi-equivalent.
\par
When $\varphi$ is a state without split property  $\varphi$ and $\varphi \circ \Theta_{R}$ 
may not be mutually quasi-equivalent.
For example, the (unique) infinite volume ground state of 
the massless XY model with spin 1/2 (d=2) gives rise to such non-equivalence.
\begin{pro} 
Let $\varphi$ be the unique infinite volume ground state of 
the massless XY model with the following Hamiltonian $H$:
\begin{equation}
H =  - \sum_{j \in {\bf Z}} \{ \sigma_{x}^{(j)}\sigma_{x}^{(j+1)} +\sigma_{y}^{(j)}\sigma_{y}^{(j+1)} \}
\label{eqn:e2}
\end{equation}
where $\sigma_{x}^{(j)}$ and $\sigma_{y}^{(j)}$ are Pauli spin matrices at the site $j$ in
one-dimensional integer lattice $\bfZ$.
\par
Then,   $\Theta_{R}$  cannot be weakly inner in the sense specified in ( \ref{eqn:e1}) .
In other words, the representations of $\frak A$ associated with 
$\varphi$ and $\varphi \circ \Theta_{R}$ are disjoint.
\label{pro:XY1}
\end{pro}
 \bigskip
 \noindent
 \par
 Does non-split property of a translationally invariant pure state imply
impossibility of obtaining a representation of the Cuntz algebra implementing $\Theta_{R}$
 on $\frak A$?
At the moment  we are not able to prove it. 
 For the proof of  Haag duality we do not need an answer to this question, though , we have 
to keep Proposition \ref{pro:XY1} in mind. 
 \bigskip
 \noindent
 \newline
 {\it Sketch of Proof \ref{pro:XY1}.}
 \\
 The XY model is formally equivalent to the free Fermion on the one-dimensional
 lattice $\bfZ$.
Our proof of Proposition \ref{pro:XY1} relies deeply on $C^*$ algebraic methods of \cite{XY2} and 
results on quasifree states of CAR algebras. As these topics are not related to the proof of 
Haag duality we present here only a sketch of proof of Proposition \ref{pro:XY1}.
\par
Let $c_{j}$ and $c^{*}_{j}$ be the creation annihilation operators of Fermions on $\bfZ$
satisfying  Canonical Anti-Commutation Relations (CAR), $\{ c_{j} , c^{*}_{k}\} =\delta_{jk}$ etc.
For $f = f(j)$ in $l^{2}(\bfZ )$ we set 
$c^{*}(f) = \sum_{j\in \bfZ} c^{*}_{j} f_{j}$ and $c(f) = (c^{*}(f))^{*}$.
By $\frak A_{CAR}$ we denote the $C^{*}$-algebra generated by $c_{j}$ and $c^{*}_{k}$.
We introduce the parity automorphism $\Theta_{parity}$ of $\frak A_{CAR}$  and the spin algebra
$\frak A$  determined by  $\Theta_{parity}(c_{j}) = - c_{j}$  and 
$\Theta_{parity}(\sigma_{x,y}^{(j)})=- \sigma_{x,y}^{(j)}$.
We set 
$$   \frak A_{CAR}^{\pm} =\{ Q \in \frak A_{CAR} | \Theta_{parity}(Q) = \pm Q \} , \quad
 \frak A^{\pm} =\{ Q \in \frak A | \Theta_{parity}(Q) = \pm Q \} .$$
 
A gauge invariant quasifree state $\psi$ of $\frak A_{CAR}$ is determined by the covariance 
operator $A$ defined by $\psi (c^{*}(f) c(g)) = ( g, Af )_{l^{2}(\bfZ)}$
where the right-hand side is the inner product of $l^{2}(\bfZ)$. 
Any bounded selfadjoint operator $A$
on $l^{2}(\bfZ)$ satisfying $0 \leq A\leq 1 $ gives rise to a quasifree state in this way,
so  by  $\psi_{A}$ we denote the gauge invariant quasifree state of $\frak A_{CAR}$ determined by 
$$  \psi_{A} (c^{*}(f) c(g)) = ( g, Af )_{l^{2}(\bfZ)}.$$
Via Jordan-Wigner transformation and $\bfZ_{2}$ cross product, Pauli spin matrices
(on $\bfZ$) are written in  terms of $c_{j}$ and $c^{*}_{j}$ and 
$ \frak A_{CAR}^{+} =  \frak A^{+}$ . 
The infinite volume ground state
$\varphi$ of  the XY model (\ref{eqn:e2}) is $\Theta_{parity}$ invariant and is determined by
 a quasifree state $\psi_{p}$ of $\frak A_{CAR}$:
$$\varphi |_{ \frak A^{+}} = \psi_{p} .$$ 
In this formula, with help of  Fourier series, $l^{2}(\bfZ )$ is identified with $L^{2}([-\pi ,\pi])$ and
 $p$ is the multiplication operator of the characteristic function $\chi_{[0,\pi]}$.

To show that   $\Theta_{R}$  is not weakly inner on the GNS space of the ground state
$\varphi$ of  the XY model, it suffices to show that $\varphi$ and $\varphi\circ\Theta_{R}$ 
are not quasi-equivalent.
To prove this claim, we focus our attention to the representation  of $\frak A_{CAR}^{+}$.
The representation of  $\frak A_{CAR}^{+}$ on the GNS space associated with
$\varphi$ has decomposition into two components , both of which are irreducible.

Now look at 
$$\varphi \circ \Theta_{R} |_{ \frak A^{+}} = \psi_{u^{*}pu}  |_{ \frak A^{+}}$$
where $u$ is an isometry on $l^{2}(\bfZ )$.
On $L^{2} ([-\pi ,\pi ] )$, $u^{*}pu$ is an operator with  a kernel function.
If $\varphi \circ \Theta_{R}$ and $\varphi $  both restricted to $\frak A^{+}$
are quasi-equivalent, the quasifree states  $\psi_{p}$ and $\psi_{u^{*}pu} $
of the CAR $\frak A_{CAR}$ must be quasi-equivalent.
(See the argument  on the top of page 99 in \cite{Voiculescu}.)
So $p - (u^{*}pu)^{1/2}$ and $(1-p) - (1- u^{*}pu)^{1/2}$ are of Hilbert Schmidt class.
These conditions imply that  $X = p - u^{*}pu$ is a Hilbert Schmidt operator.
 However, the kernel $k(\theta_1, \theta_2 )$ for the operator $X$ has a singularity
 of order $|\theta_1 - \theta_2 |^{-2}$ at the diagonal part. Thus 
\begin{equation}
 {\rm Tr} (X^*X) =  {\rm Tr}(( p - u^{*}pu)^2) = \infty .
\label{eqn:e3}
\end{equation}
Thus (\ref{eqn:e3}) leads a contradiction if  $\varphi$ and $\varphi\circ\Theta_{R}$ 
are quasi-equivalent.
\newline
{\it End of Sketch of Proof \ref{pro:XY1}.}
 
\newpage
\section{$O_{d}\otimes O_{d}$}\label{SecCuntz}
\setcounter{theorem}{0}
\setcounter{equation}{0}
Our basic strategy to prove Theorem \ref{th:duality} is the following.
We consider the gauge invariant extension $\overline{\psi}$ of 
the state $\varphi$ to the Cuntz algebra $O_{d}\otimes O_{d}$ and 
examine conditions of factoriality of $\varphi$. Then, we consider a pure state $\psi$ of $O_{d}\otimes O_{d}$ which is a pure state extension of $\varphi$ and prove Haag duality 
at the level of the Cuntz algebra.
\\
\\
Next we introduce our notation for the Cuntz algebra $O_d$.
The Cuntz algebra $O_d$ is a simple $C^*$-algebra generated by isometries 
$S_{1} , S_{2} \cdots S_{d}$ satisfying 
$S_k^* S_l \: = \: \delta_{kl} 1$ , $\sum_{k=1}^{d} \: S_k S_k^* \:  = \: 1$.
The gauge action $\gamma_{U}$ of the group $U(d)$ of d by d unitary matrices is 
defined via the following formula:
$$\gamma_{U}(S_k) \; = \; \sum_{l=1}^{d} U_{lk}S_l .$$
where $U_{kl}$ is the k l matrix element  for $U$ in $U(d)$.
Consider the diagonal circle group $U(1)= \{ z \in   {\bf C} | \abs{z} = 1 \}$ 
and $\gamma_z$ on $O_{d}$, $\gamma_z( S_j ) = z S_j $, ( $ j = 1,2,\cdots d $).
The fixed point algebra ${O_d}^{U(1)}$ for this action of $U(1)$
is the UHF algebra $d_{\infty}$ which we will identify with 
${\frak A}_R={\frak A}_{[1,\infty)}$ as follows:
Let $I$ and $J$ be m-tuples of ordered indices,
$ I = (i_1, i_2, i_3 , \cdots, i_m)$,  $J = (j_1, j_2, j_3 , \cdots , j_m)$ ($i_k , j_l \in \{1,2, \cdots , d\} $) 
and set
$S_I = S_{i_1} S_{i_2} \cdots S_{i_m} $, $S_J = S_{j_1} S_{j_2}\cdots S_{j_m} $.
Then, we identify the matrix unit of ${\frak A}_R$ and 
the $U(1)$ gauge invariant part of $O_d$ via the following equation:
$$ S_I  S_J^* = e_{i_1 j_1}^{(1)} e_{i_2 j_2}^{(2)}.....e_{i_m j_m}^{(m)}$$
where $e_{ij}$ is the matrix unit of the one-site matrix algebra.
The canonical endomorphism  $\Theta$ of ${O_d}$ is determined by
$$\Theta (Q) =   \sum_{k=1}^{d}  S_k Q S_k^*  \quad\quad Q \in O_d .$$
It is easy to see that the restriction of  $\Theta$ to $\frak A_R$ is the lattice translation $\tau_1$. 
\\
\\
\begin{lmm}
Let $\varphi$ be a translationally invariant factor state of  ${\frak A}$.
Consider the restriction   $\varphi_{R}$ of $\varphi$ to  ${\frak A}_R$.
Let $\tilde{\psi}$ be the $U(1)$ gauge invariant extension of $\varphi_R$ to $O_d$. 
Suppose further that $\tilde{\psi}$ is not factor.
\\
Then, there exists a positive $k$ such that $\tau_{k}$ acting on  ${\frak A}_R$ is 
weakly inner on the GNS spaces associated with $\varphi$ and $\tilde{\psi}$.
 More precisely, there exists a representation $\tilde{\pi}(\cdot )$ of the Cuntz algebra $O_{d\times k}$
 on the GNS space ${\frak H}_{\tilde{\psi}}$ such that
 \begin{equation}
 \tilde{\pi}(S_{l}) \in \pi_{\tilde{\psi}}({\frak A}_{R})^{\prime\prime} , \quad
 \sum_{l=1}^{dk}  
 \tilde{\pi}(S_{l}) \pi_{\tilde{\psi}}(Q) \tilde{\pi}(S_{l}^{*}) =  \pi_{\tilde{\psi}}(\tau_{k}(Q)) .
 \label{eqn:b2}
 \end{equation}
\label{lmm:extension1}
 $\tilde{\pi}(S_{l})$ implements the canonical endomorphism of $O_{d\times k} $ as well. 

Conversely, if there exist operators $T_j$ in $\pi_{\tilde{\psi}}({\frak A}_{R})^{\prime\prime}$
satisfying 
$$ \sum_{j=1}^{d}  T_j \pi_{\tilde{\psi}}(Q) T_j^{*} =  \pi_{\tilde{\psi}}(\Theta(Q)) $$
for any $Q$ in $O_d$,  $\tilde{\psi}$ is not a factor.
\end{lmm}
{\it Proof.} Let $\tilde{\psi}$ be the $U(1)$ gauge invariant extension of $\varphi$
to $O_{d}$ and $\{ \pi(O_{d}), \Omega , \frak H\}$ be the GNS representation 
associated with $\tilde{\psi}$. ($\Omega$ is the GNS cyclic vector.)
There exists a unitary representation $U_{z}$ of $U(1)$ satisfying 
\begin{equation}
 U_{z} \Omega = \Omega , \quad\quad U_{z}\pi (Q) U_{z}^{*} = \pi (\gamma_{z} (Q)) 
\quad\quad \mbox{ for $Q \in O_{d}$.}
\label{eqn:b2000}
 \end{equation}
 We set $\calN = \pi (O_{d})^{\dprime}$ and $\calC = \calN \cap \calN^{\prime}$.
Using $U_{z}$ we have introduced the normal extension $\gamma_{z}$ of U(1) action to 
the von Neumann algebra $\calN$. 
(By abuse of notation we use the same symbol $\gamma_{z}$ for this action.)
Let $Q$ be an element of  $\calN = \pi (O_{d})^{\dprime}$. 
and consider Fourier expansion of $Q$:
\begin{equation}
 Q = \sum_{k =-\infty}^{\infty} Q_{k} , \quad Q_{k} = \int dz  z^{-k} U_{z} Q U_{z}^{*} 
 \label{eqn:b2001}
 \end{equation}
Let $\calN_{k}$ be the subspace generated by operators $Q_{k}$:
\begin{equation}
\calN_{k} = \{ Q \in \calN  \: | \: \gamma_{z}(Q) = z^{k} Q \}, \quad \calN_{0} = \pi (\frak A_{R})^{\dprime}. 
\label{eqn:b2002}
 \end{equation}
 Let
 $$\calC_{k} = \calN_{k} \cap \calC   = \{ Q \in \calC  \: | \: \gamma_{z}(Q) = z^{k} Q \}. $$
As we assumed that $\calN$ is not a factor, we can find a non-trivial self-adjoint  element  $c$ of 
the center $\calC$.
As $\calN_{0}$ is a factor on $\overline{\calN_{0}\Omega}$ and $\Omega$ is cyclic for $\calN$,
 $c_{0}$ is a scalar multiple of the identity, i.e. $c_{0} = c 1$.
 As $c_{k}c_{-k}$ and $c_{k}c_{k}^{*}$ belong to $\calC$, and
since we assume that $C$ is self-adjoint $c_{k}c_{-k} = c_{k}c_{k}^{*}$ is scalar.
By the same reason, $c_{-k}c_{k}$ and $c^{*}_{k} c_{k}$ are scalar as well.
Thus by rescaling we can assume that any non-vanishing $c_{k}$ is a unitary.
Moreover if $\calC_{k} $ is not $0$ it is one-dimensional. To see this take another central element 
$c^{1}$ and consider its Fourier component $c^{1}_{k}$. 
As $c^{1}_{k } c_{-k}$ belongs to $\calC_{0}$, it is a scalar.
\par
Take the smallest positive $k$ such that  $\calC_{0}$ is one-dimensional and for a multi-index $I$
with $| I | = k$, we set 
$$ \tilde{\pi}(S_{I})=  \pi(S_{I})c_{k}^{*} ,  \quad\quad \tilde{\pi}(S_{I}^{*} ) = \pi(S_{I}^{*}) c_{k} .$$ 
Both $\tilde{\pi}(S_{I})$ and $\tilde{\pi}(S_{I})^{*}$ are $\gamma_{z}$ invariant and
their restriction to $\frak H_{\varphi_{R}}$ satisfies (\ref{eqn:b2}).

Next let $T_j$ be a operators in $\pi_{\tilde{\psi}}({\frak A}_{R})^{\prime\prime}$
implementing the canonical endomorphism $\Theta$ of $O_d$.
Then the operator $\pi_{\tilde{\psi}}(S^*_j) T_i $ commutes with any element of 
$\pi_{\tilde{\psi}}(O_d )^{\prime\prime}$ because of 
$$\pi_{\tilde{\psi}}(S^*_j) \pi_{\tilde{\psi}}(\Theta (Q)) = 
\pi_{\tilde{\psi}}(Q) \pi_{\tilde{\psi}}(S^*_j) .$$
Thus $\pi_{\tilde{\psi}}(O_d )^{\prime\prime}$ is not a factor.
{\it End of Proof.}
\\
\\
The following lemma is known. (See, for example, Lemma 6.10 and 6.11 of \cite{BJKW}.) 
\begin{lmm}
Let $\varphi$ be a translationally invariant factor state of  ${\frak A}$.
Suppose that for a positive $k$,
the restriction $\tau_{k}$ to  ${\frak A}_R$ is implemented by a representation
$\tilde{\pi}(O_{d\times k})$
of the Cuntz algebra $O_{d\times k}$ on the GNS space ${\frak H}_{\varphi_{R}}$
and the gauge invariant part of $\tilde{\pi}(O_{d\times k})$ coincides with
${\frak A}_R$. More precisely,
$$  \tilde{\pi}(S_{l}S_{k}^{*})  =  \pi_{\varphi_{R}}(e_{kl}^{(1)}) .$$
Suppose that the gauge action $\gamma_{z}$ does not admit  a normal extension to
the von Neumann algebra $\tilde{\pi}(O_{d\times k})^{\dprime}$ for any $z$. 
Then, $\tau_{k}$ is weakly inner in the sense of (\ref{eqn:b2}), namely
$\tilde{\pi}(O_{d\times k})^{\dprime} = {\frak A}_R^{\dprime}$.
\label{lmm:WeaklyInner1}
\end{lmm}
{\it Proof} : By abuse of notation $\varphi_{R}$ is regarded as a state of the fixed
point subalgebra $(O_{d\times k})^{U(1)}$.
Consider  a vector state  $\psi_0$ of $(O_{d\times k})^{U(1)}$
associated with the GNS vector for $\varphi_R$ and let 
  $\psi$ be the $U(1)$ invariant extension of $\varphi_{R}$ to $(O_{d\times k})^{U(1)}$. 
 Then, $\int \psi_0\circ \gamma_z  dz = \psi$ and at the level of the GNS representation,
$$\frak H_{\psi}  =  \int^{\otimes} \frak H_{\psi_0} \: \: dz
= \frak H_{\psi_0} \otimes L^2 (S^1)  
, \quad 
\pi_{\psi} = \int^{\otimes} \pi_{\psi_0}\circ \gamma_z \: \: dz$$  
Due to our assumption that $\gamma_z$ does not admit any normal extension to 
$\tilde{\pi}(O_{d\times k})^{\dprime}$ for any $z$,
the von Neumann algebra $\calN = \pi_{\psi} (O_{d\times k})^{\dprime}$ is isomorphic
to ${\frak M}\otimes L^{\infty}(S^{1})$ where the gauge action acts as the rotation on $S^{1}$.
$\pi_{\psi} ({\frak A}_{R})^{\dprime}$ is the commutant of the unitaries implementing the rotation.
\begin{equation}
 \pi_{\psi} ({\frak A}_{R})^{\dprime} = {\frak M}\otimes 1
 \label{eqn:b301}
 \end{equation}
By definition,  $\pi_{\psi}(Q) =\pi_{\psi_0}(Q)\otimes 1$ for $Q$ in  ${\frak A}_{R}$
and we have
\begin{equation}
 \pi_{\psi} ({\frak A}_{R})^{\dprime} = \pi_{\varphi_{R}} ({\frak A}_{R})^{\dprime} \otimes 1
\label{eqn:b302}
 \end{equation}
Looking at each fiber of equations (\ref{eqn:b301}) and (\ref{eqn:b302}) , we conclude that 
$\tilde{\pi}(O_{d\times k})^{\dprime} = {\frak A}_R^{\dprime}$.
{\it End of Proof.}
\\
\\
\par
Next we consider a pair of Cuntz algebras denoted by $O_d^{(L)}$ and $O_d^{(R)}$
and we set $\calB =O_{d}^{(L)} \otimes O_{d}^{(R)}$.
The Cuntz generators are denoted by $S_I^{(L)}$ and  $S_I^{(R)}$ etc. 
The algebra $\calB$ is naturally equipped with  the $U(1)\otimes U(1)$ gauge action 
$\gamma_{z_L ,z_R} =\gamma_{z_L}\otimes \gamma_{z_R}$ :
$$\gamma_{z_L ,z_R}(S_I^{(L)}) = z_L^{| I |}  S_I^{(L)} , \quad 
\gamma_{z_L ,z_R}(S_I^{(R)})= z_R^{| I |}  S_I^{(R)}   \quad ( z_L ,z_R \in U(1) ) $$
As ${\frak A} = {\frak A}_{L}\otimes{\frak A}_{R}$ we identify
${\frak A}$ with the $U(1)\otimes U(1)$ fixed point sub-algebra 
$\calB =O_{d}^{L}\otimes O_{d}^{R}$.
The canonical endomorphisms of $\calB$ is defined via the following equation: 
$$ \Theta_{k,l} = \Theta_{L}^{k}\otimes\Theta_{R}^{l}$$
where $\Theta_{L}$ (resp. $O_{d}^{R}$) is the canonical endomorphism of $O_d^{(L)}$
(resp. $O_d^{(R)}$).
\\
\\
\par
The lattice translation automorphism $\tau_{1}$ has an extension to $\calB$ as an inner automorphism. 
To see this, set   
\begin{equation}
 V = \sum_{j=1}^{d}  (S_j^{(L)})^{*} S_j^{(R)} . 
 \label{eqn:b3}
 \end{equation}
Then, $ V$ satisfies
 \begin{equation}
 V V^{*} = V^{*} V = 1 , \quad 
V e_{kl}^{(0)} V^{*} =  V  S_k^{(L)} (S_l^{(L)})^{*}V^{*} =
S_k^{(R)} (S_l^{(R)})^{*} = e_{kl}^{(1)}
 \label{eqn:b4}
 \end{equation}
which shows that 
\begin{equation}
Ad (V) (Q) = \tau_{1} (Q)  \quad\quad Q \in {\frak A}
 \label{eqn:b5}
 \end{equation}
We extend $\tau_{1}$ to $\calB$ via the above equation (\ref{eqn:b5}).
\\
\\
Let $k$ be a positive integer and we regard $O_{d\times k}$ is a subalgebra of $O_{d}$
which is generated by $S_{I}$ and   $S_{J}^{*}$ with $| I | = k n$ ,$ |J|= km $ ($n,m=1,2,\cdots$ ).
Set 
$$\calB^{k} =O_{d\times k}^{L} \otimes O_{d\times k}^{R} \subset \calB.$$
\begin{lmm}
Let $\varphi$ be a pure state of  ${\frak A}$.
Suppose that there exists a representation $\tilde{\pi}$ of $\calB^{k}$
on the GNS space $\frak H_{\varphi}$ associated with $\varphi$
such that 
\begin{eqnarray}
&&\tilde{\pi}(S_I^{(L)} (S_J^{(L)})^{*}) = 
\pi_{\varphi}(e_{i_{1}j_{1}}^{(0)} e_{i_{2}j_{2}}^{(-1)}
\cdots e_{i_{k}j_{k}}^{(-k+1)}) ,
\nonumber\\
&&\tilde{\pi}(S_I^{(R)} (S_J^{(R)})^{*}) =\pi_{\varphi}(e_{ij}^{(1)}
e_{i_{1}j_{1}}^{(2)}  \cdots e_{i_{k}j_{k}}^{(k)}) . 
 \label{eqn:b6}
 \end{eqnarray}
Then,  $\tilde{\pi}(O_{d\times k}^{(L)}) \subset \pi_{\varphi}({\frak A}_{L})^{\dprime}$ and
$\tilde{\pi}(O_{d\times k}^{(R)} ) \subset \pi_{\varphi}({\frak A}_{R})^{\dprime}$. 
\label{lmm:WeaklyInner2}
\end{lmm}
{\it Proof}:  Due to Lemma \ref{lmm:WeaklyInner1}, we have only to show the gauge action
$\gamma$ does not have a normal extension to the von Neumann algebra 
$\tilde{\pi}(\calB )^{\dprime}$.
Any normal homorophism of a type I factor is implemented by a unitary.
As $\tilde{\pi}(\calB )$ is irreducible, we suppose there exists a unitary $W$ such that 
\begin{equation}
W \tilde{\pi}( S_i^{(L)} ) W^{*} =    z_{L} S_i^{(L)} , \quad       
W \tilde{\pi}( S_i^{(R)} ) W^{*} =    z_{R} S_i^{(R)}.
 \label{eqn:b601}
 \end{equation}
Then, due to  (\ref{eqn:b6})
$W$ commutes with the gauge invariant part $\calA_{L}$ and $\calA_{R}$.
As $\varphi$ is pure, $W$ is a scalar multiple of the identity and  $z_{L} =z_{R}= 1$. 
\\
{\it End of Proof}
\\
\\
\begin{lmm}
Let $\varphi$ be a translationally invariant pure state of  ${\frak A}$ and
let $\overline{\psi}$ be the $U(1)\times U(1)$ gauge invariant extension
of $\varphi$ to $\calB$.
\begin{equation}
\overline{\psi} (Q) = \varphi \left( \int_{U(1)\times U(1)}
 \gamma_{z_{L} z_{R}} (Q) \:\:  dz_{L} dz_{R} \right) \quad\quad Q \in \calB .
 \label{eqn:b7}
 \end{equation}
  $\overline{\psi}$ is not a pure state.
\label{lmm:extension-A}
\end{lmm}
{\it Proof} :  Let $\{ \overline{\pi}(\calB ), \Omega,  \overline{\frak H} \}$ be the GNS triple.
As $\overline{\psi}$ is  $\gamma_{z_{L} z_{R}}$ invariant, there exists a unitary $U_{z_{L} z_{R}}$
satisfying 
$$U_{z_{L} z_{R}}  \overline{\pi}(Q) U_{z_{L} z_{R}}^{*} =   \overline{\pi}(\gamma_{z_{L} z_{R}}(Q))
, \quad U_{z_{L} z_{R}}\Omega = \Omega$$
We consider the restriction of $\overline{\pi}$ to $\frak A$ and
the Fourier decomposition of  $\overline{\frak H}$ with respect to $U_{z_{L} z_{R}}$ .
$$ \overline{\frak H} = \sum_{k,l \in \bfZ} \:\: \oplus \:\: \overline{\frak H}_{kl} .$$
If $\overline{\psi}$ is pure, 
$$ \overline{\pi}(\frak A)^{\dprime} = \overline{\pi}(\calB)^{\dprime}\cap \calC^{\prime} = \calC^{\prime} $$
where $\calC$ is   the abelian von Neumann algebra generated by $U_{z_{L} z_{R}}$.
As $\overline{\pi}(\frak A)^{\dprime}$ is the commutant of $\calC$,
the center of $\overline{\pi}(\frak A)^{\dprime}$ is $\calC$. 
Each irreducible representation $\pi (\frak A)$ appearing 
in $\overline{\pi} (\frak A)$ as a subrepresentation is 
of the form $\overline{\pi}(Q)P$ where $P$ is a central projection of  $\overline{\pi}(\frak A)^{\dprime}$.
 Thus $\overline{\pi}(\frak A)^{\dprime}$
is decomposed into irreducible representations $\pi_{kl}$ on $\overline{\frak H}_{kl} $. 
$\pi_{kl}$ and $\pi_{nm}$ are equivalent if and only if $k=n$,and $l=m$. 
$\pi_{00}$ is equivalent to the GNS representation associated with $\varphi$. 
However the operator $\overline\pi (V)$ gives rise to unitary equivalence between
$\pi_{00}$ and $\pi_{1-1}$ , which implies contradiction.
 Thus $\overline{\psi}$ cannot be pure.
{\it End of Proof.}
\\
\par
By the same line of argument in Lemma \ref{lmm:extension1}, we can show that
the Fourier component $\calC_{ij}$ of $\calC$ in Lemma \ref{lmm:extension-A}, 
is either one or zero dimensional.  Furthermore  $\calC$ is generated by $\calC_{k ,-k}$ for some  $k$ 
when the canonical endomorphism is not weakly inner in $\pi_{\varphi}(\frak A)^{\dprime}$.
We show this claim rather implicitly in the next step.
\\
\par
We introduce the diagonal action $\gamma^{d}_{z}$ of $U(1)$  on $\calB$ via the equation:
$\gamma^{d}_{z} = \gamma_{z , z}$ and similarly
the diagonal action $\gamma^{d,k}_{z}$ of $U(1)$  on $\calB^{k}$ 
Set
$$\calD = \{ Q \in \calB \:\: | \:\: \gamma^{d}_{z}(Q) = Q \quad\mbox{for any $z$.} \}$$
\begin{lmm}
(i) $\calD$ is generated by $\frak A$ and $V$, hence    $\calD$ is isomorphic to the crossed
product of ${\frak A}$ by the action $\tau_{j}$ of $\bfZ$.
\\
(ii) Let $\varphi$ be a translationally invariant state of $\frak A$. There exists a state
$\tilde{\varphi}$ of $\calD$ satisfying   
\begin{equation}
\tilde{\varphi}(V) = 1 ,  \quad\quad \tilde{\varphi} (Q) = \varphi (Q)  \quad Q \in {\frak A} .
\label{eqn:b8}
\end{equation}
The state $\tilde{\varphi}$ of $\calD$ satisfying (\ref{eqn:b8}) is unique.
\\
(iii) $\tilde{\varphi}$ is pure if  $\varphi$ is factor.
\label{lmm:extension-B}
\end{lmm}
{\it Proof}:
(i) $\calD$ is generated by  $S_{I}^{(L)}(S_{J}^{(R)})^{*} Q$
where multi-indices $I$ and $J$ satisfy $| I | - |J| = 0$ and  $Q$ is  an element of $\frak A$.
By direct calculation, we have $V S_{i}^{(L)}(S_{j}^{(R)})^{*}  =  S_{i}^{(R)}(S_{j}^{(R)})^{*}$.
Thus
$$S_{i}^{(L)}(S_{j}^{(R)})^{*} Q = V^{*} VS_{i}^{(L)}(S_{j}^{(R)})^{*} Q = 
V^{*} S_{i}^{(R)}(S_{j}^{(R)})^{*}Q$$
which shows that
 $S_{I}^{(L)}(S_{J}^{(R)})^{*} Q$ is written by a product of $V$ and elements in   $\frak A$.
\\
(ii) Consider the GNS triple $\{ \pi_{\varphi}(\frak A) , \Omega , {\frak H}_{\varphi}\}$ associated with
$\varphi$. As the state $\varphi$ is translationally invariant  we have a unitary $W$ implementing
$\tau_{1}$ and $W\Omega = \Omega$. Then we set $\pi_{\varphi}(V) =W$ the vector state
$\tilde{\varphi}$ of $\calD$ associated with $\Omega$ satisfies (\ref{eqn:b8}). 
Conversely, if a state $\tilde{\varphi}$ satisfies (\ref{eqn:b8}), the GNS cyclic vector
$\Omega_{\tilde{\varphi}}$ is invariant under $\pi_{\tilde{\varphi}}(V)$ due to the identity:
$$||( \pi_{\tilde{\varphi}}(V)- 1)\Omega_{\tilde{\varphi}}||^{2} = 2 - \tilde{\varphi}(V)-\tilde{\varphi}(V^{*})
=2-1-1=0 .$$
Thus $W=\pi_{\tilde{\varphi}}(V)$.
\\
(iii) As $\varphi$ is factor, for $Q \in {\frak A}$
$$w-\lim_{k \to \infty}  \pi_{\varphi}(\tau_{k}(Q)) = \varphi (Q) 1.$$
Suppose $P$ commutes with $\pi_{\tilde{\varphi}}(V)$ and $\pi_{\varphi}({\frak A})$.
Then,
\begin{eqnarray}
&&(\Omega , \pi_{\varphi} (Q) P\Omega) = 
(\Omega , \pi_{\varphi} (Q) P \pi_{\tilde{\varphi}}(V^{-k})\Omega)
=(\Omega , \pi_{\varphi} (Q) \pi_{\tilde{\varphi}}(V^{-k}) P\Omega)
\nonumber\\
&=&(\Omega , \pi_{\varphi} (Q) \pi_{\tilde{\varphi}}(V^{-k}) P\Omega)
= (\Omega , \pi_{\varphi} (\tau_{k}(Q)) P\Omega)
 \nonumber\\
&=&\lim_{k \to \infty}(\Omega , \pi_{\varphi} (\tau_{k}(Q)) P\Omega) 
= \varphi (Q) (\Omega , P\Omega) 
\label{eqn:b801}
\end{eqnarray}
which implies 
$P\Omega =  (\Omega , P\Omega)\Omega$, $P = (\Omega , P\Omega) 1$.
\\
{\it End of Proof.}
\\
\begin{lmm}
Let $\varphi$ be a translationally invariant pure state of  ${\frak A}$. Then, for
a positive $k$ there exists a pure state extension  $\psi$ of $\varphi$ to $\calB^{k}$ 
such that $\psi$ is invariant under  $\tau_{k}$ and
\begin{equation}
\sum_{| I | =k} \psi ( (S_{I}^{(L)})^{*} S_{I}^{(R)}) = 1.
 \label{eqn:b9}
 \end{equation}
Furthermore, one of the following mutually exclusive conditions is valid. 
 \\
 (i) $\psi$ is invariant under $\gamma^{d,k}_{z}$ .
 \\
 (ii) $\psi \circ\gamma^{d,k}_{z}$ is not equivalent to $\psi$ for any $z$.
 \\
 \\
 When (ii) is valid, the assumptions of Lemma \ref{lmm:WeaklyInner2}
are satisfied.
\label{lmm:extension-C}
\end{lmm}
{\it Proof}: Consider the state $\tilde{\varphi}$ of $\calD$ satisfying (\ref{eqn:b8}). 
Let $\tilde{\psi}$ be the $\gamma^{d}$ invariant extension of $\tilde{\varphi}$ to $\calB$.

If $\tilde{\psi}$ is pure, we set $\tilde{\psi} = \psi$ and 
as $\varphi$ is translationally invariant, there exists
a unitary $W$ on ${\frak H}_{\varphi} ={\frak H}_{0}$ satisfying
$$ W \pi_{\varphi} (Q) W^{*} = \pi (\tau_{1}(Q)) , \quad W\Omega_{\varphi} =\Omega_{\varphi} .$$ 
Then the operator $\pi_{\tilde{\varphi}}(V)  W^{*}$ acting on $\frak H_{\varphi}$ 
commutes with $\pi_{\varphi} (\frak A)$ . 
This shows that  $\pi_{\tilde{\varphi}}(V) W^{*}$  is a scalar.
 After a gauge transformation of $O_{d}^{(L)}$ we have
$$\pi_{\tilde{\varphi}}(V) \Omega_{\tilde{\varphi}} = \Omega_{\tilde{\varphi}}$$ 
which is equivalent to the equation (\ref{eqn:b9}). By definition the state
$\psi$ is $\gamma^{d,k}_{z}$ invariant.

Next we consider the case that $\tilde{\psi}$ is not pure.
Let $U(z)$ be the unitary on the GNS space ${\frak H}_{\tilde{\psi}}$ 
associated with $\tilde{\psi}$ such that 
$$\pi_{\tilde{\psi}}(\gamma^{d}_{z}(Q)) = U(z)\pi_{\tilde{\psi}}(Q) U(z)^{*} ,
\quad  U(z)\Omega_{\tilde{\psi}} = \Omega_{\tilde{\psi}} .$$
The GNS representation $\pi_{\tilde{\psi}}$ restricted to $\calD$ is a direct sum of 
$\pi_{j}(\calD)$ on ${\frak H}_{j}$:
$${\frak H}_{\tilde{\psi}} = \sum_{j \in \bfZ} \: {\frak H}_{j} ,  \quad \quad U(z) |_{ {\frak H}_{j}} = z^{j} 1$$ 
$$\pi_{\tilde{\psi}} =  \sum_{j \in \bfZ} \oplus \pi_{j} , \quad \quad  
 \pi_{j} =  \pi_{\tilde{\psi}}|_{{\frak H}_{j} } $$
 Note that the representations $\pi_{j}$ and $\pi_{i}$ are disjoint when $i\ne j$.

The Fourier component of the commutant $\calC$ of $\pi_{\tilde{\psi}}(\calB)$ 
is denoted by $\calC_{j}$. 
For $Q$ in $\calC =  \pi_{\tilde{\psi}}(\calB)^{\prime}$
$$ Q = \sum_{k =-\infty}^{\infty} Q_{k} , \quad Q_{k} = \int dz  z^{-k} U_{z} Q U_{z}^{*} $$
Let $\calC_{k}$ be the subspace generated by operators $Q_{k}$:
$$\calC_{k} = \{ Q \in \calC  |  \gamma_{z}(Q) = z^{k} Q \}$$
As the state $\varphi$ is pure $\calC_{0}$ is one dimensional,
$\calC_{0}= {\bf C}1$  because $\calC_{0}$ commutes with $\pi_{\varphi}(\frak A )$.
By the similar argument in proof of  Lemma \ref{lmm:extension1},
it is possible to show  the dimension of $\calC_{k}$
is zero or one and $\calC$ is generated by a single unitary $U$ in 
$\calC_{k}$ for some $k$.

Now we introduce a representation $\pi (\calB_{k})$ of $\calB_{k}$ on 
${\frak H}_{0} = {\frak H}_{\varphi}$ determined by 
$$\pi ( S_{I}^{(L)}) =  e^{i\theta} \pi_{\tilde{\psi}}( S_{I}^{(L)}) U^{*} |_{{\frak H}_{0}} , \quad
\pi ( S_{J}^{(R)}) = \pi_{\tilde{\psi}}( S_{J}^{(R))}) U^{*} |_{{\frak H}_{0}} \quad
\mbox{for $ | I | = | J | = k$}$$
where the phase factor $e^{i\theta}$ is determined later.
By definition $\pi (Q)= \pi_{\varphi}(Q)$ for $Q$ in $\frak A$ while
on ${\frak H}_{0}$, $\pi_{\varphi}(\frak A)$ acts irreducibly.
Let $\psi$ be the vector state of $\calB^{k}$ associated with
$\Omega_{\varphi}$. As $\varphi$ is translationally invariant, there exists
a unitary $W$ on ${\frak H}_{\varphi} ={\frak H}_{0}$ satisfying
$$ W \pi_{\varphi} (Q) W^{*} = \pi (\tau_{k}(Q)) , \quad W\Omega_{\varphi} =\Omega_{\varphi} .$$ 
Set 
$$V^{(k)} = \sum_{| I | =k}  (S_{I}^{(L)})^{*} S_{I}^{(R)} .$$
Then the operator $V^{k}W^{*}$ commutes with $\pi_{\varphi} (\frak A)$ . This shows
that  $\pi (V^{(k)})W^{*}$  is a scalar. By suitably choosing  the phase factor $e^{i\theta}$
we have 
$$\pi (V^{(k)} )=W , \quad \pi (V^{(k)})\Omega_{\varphi} =\Omega_{\varphi} .$$ 
$\psi$ is the state satisfying our requirement.
{\it End of Proof.}
\newpage
\section{Proof of Theorem \ref{th:duality}}\label{Proof}
\setcounter{theorem}{0}
\setcounter{equation}{0}
We consider Haag duality for the Cuntz algebras $O_{d} \otimes O_{d}$ first. 
The same duality (Proposition  \ref{pro:CuntzDual2}) is stated  in \cite{BJKW}.
 However, due to the reason stated in the introduction of this paper, we present our proof here.
To show the Haag duality for the Cuntz algebra we apply Tomita-Takesaki Theory.
The state $\varphi_{R}$ or its extension
to $O_{d}$ may not be faithful so  we consider reduction of
the von Neumann algebra generated by $O_{d}$  by support projection and
apply the Tomita modular conjugation to obtain the (reduced) commutant.
Then we apply the following lemma.
\\
\begin{lmm}
Let $\calM_{1} \subset \calM_{2}$ be a pair of factor-subfactor on a separable 
Hilbert space $\mathfrak H$.
Suppose  that there exists a projection
$P$ in $\calM_{1}$ such that  $P\calM_{1}P=P\calM_{2}P$. Then,  $\calM_{1}$ and $\calM_{2}$
coincide:  $\calM_{1}=\calM_{2}$.
\label{lmm:lifting2} 
\end{lmm}
{\it Proof.}
Suppose that  we have a matrix unit $e_{ij}$ ($i,j =1,2, \cdots.$) in $\calM_{1}$ such that
\begin{equation}
e_{11}=P , \quad \sum_{j=1} e_{jj} =1
\label{eqn:b1001}
\end{equation}
Let $Q$ be an element of  $\calM_{2}$. Then $e_{ii} Q e_{jj}$ is an element of
 $\calM_{1}$ because 
 $$e_{ii} Q e_{jj}= e_{i1}e_{1i} Q e_{j1} e_{1j} ,\quad e_{1i} Q e_{j1} \in \calM_{1} .$$
Thus if we have a matrix unit satisfying (\ref{eqn:b1001}) any $Q$ in  $\calM_{2}$
 is an element of  $\calM_{1}$.
When $\calM_{1}$ has a tracial state $tr$ and $1/ tr P$ is not an integer, the matrix unit
satisfying $\sum_{j=1} e_{jj} =1$ does not exists. In such a case, we consider another projection
$q$ in $\calM_{1}$ such that $ q \leq P$ and  $1/ tr q$ is a positive integer.
Then we apply the above argument to $qM_{1}q = qM_{2}q$.
{\it End of Proof.}
\bigskip
\bigskip
\bigskip
\noindent
\par
Without loss of generality, we assume that $k=1$ in Lemma \ref{lmm:extension-C}
for the proof of Haag duality.
Let $\varphi$ be a translationally invariant state. 
From now on, $\psi$ is the pure state extension of $\varphi$ to $\calB$ such that  
$\psi$ is invariant under  $\tau_{1}$.
Recall that due to the equation (\ref{eqn:b9}) ,
$$ \pi_{\psi} (V) \Omega_{\psi} =  \Omega_{\psi}.$$
Hence, 
$$(S_{I}^{(R)})^{*} V =  (S_{I}^{(L)})^{*} , \quad
 \pi_{\psi} (S_{I}^{(R)})^{*} \Omega_{\psi} = \pi_{\psi} (S_{I}^{(R)})^{*} \pi_{\psi} (V) \Omega_{\psi} = \pi_{\psi} (S_{I}^{(L)})^{*} \Omega_{\psi}. $$
\begin{equation}
\pi_{\psi} (S_{I}^{(L)})^{*} \Omega_{\psi} = \pi_{\psi} (S_{I}^{(R)})^{*} \Omega_{\psi}. 
 \label{eqn:b10}
 \end{equation}
 As a consequence the Hilbert space $\frak H_{\psi}$ is generated by the following vectors: 
 \begin{equation}
 \pi_{\psi} (S_{I}^{(L)})  \pi_{\psi} (S_{J}^{(R)})\pi_{\psi}(Q) \pi_{\psi} (S_{J'}^{(R)})^{*}   \Omega_{\psi}
\: , \quad Q \in {\frak A_{R}}.
\label{eqn:b11}
\end{equation}
\begin{pro} 
Suppose $\psi$ is the pure state extension of $\varphi$ to $\calB$ such that  
$\psi$ is invariant under  $\tau_{1}$. Then,
\begin{equation}
 \pi_{\psi}(O_{d}^{(L)})^{\dprime} = \pi_{\psi}(O_{d}^{(R)})^{\prime}.
\label{eqn:b12}
 \end{equation}
 \label{pro:CuntzDual2}
 \end{pro}
The equation (\ref{eqn:b10}) is crucial in our proof of (\ref{eqn:b12}).
We need some preparation for our proof of (\ref{eqn:b12}).
\par
Let  $E_{R}$ be the support projection of $\psi$ for $\pi_{\psi}(O_{d}^{(R)})^{\dprime}$
and $E_{L}$ be the support projection of $\psi$ for $\pi_{\psi}(O_{d}^{(L)})^{\dprime}$.
By $E_{R}^{\prime}$ we denote the projection with range
 $[ \pi_{\psi}(O_{d}^{(R)})^{\dprime} \Omega_{\psi}]$
where $[ \pi_{\psi}(O_{d}^{(R)})^{\dprime} \Omega_{\psi}]$ is the closed subspace of
$\frak H_{\psi}$ generated by $ \pi_{\psi}(O_{d}^{(R)}) \Omega_{\psi}$.
Similarly, by $E_{L}^{\prime}$ we denote the projection to 
$[ \pi_{\psi}(O_{d}^{(L)})^{\dprime} \Omega_{\psi}]$.
\par
Set $P = E_{R}^{\prime}E_{R}$ and $\frak K = P\frak H$. 
The range of $P$ is $[ E_{R} \pi_{\psi}(O_{d}^{(R)})^{\dprime} \Omega_{\psi}]$ .

Now we denote the von Neumann algebra $E_{R} \pi_{\psi}(O_{d}^{(R)})^{\dprime}E_{R}$
 by $\frak N$.
$\Omega_{\psi}$ is a cyclic and separating vector for $\frak N$
acting on $\frak K$. Let $\Delta$ and $J$ be the Tomita modular operator and the modular
conjugation associated with $\Omega_{\psi}$
for $E_{R} \pi_{\psi}(O_{d}^{(R)})^{\dprime}E_{R}$.
Set $v_{j} = P\pi_{\psi}(S_{j}^{(R)})P$. 
As  
$$(\Omega_{\psi}, \left((P\pi_{\psi}(S_{j}^{(R)})P - P\pi_{\psi}(S_{j}^{(R)})\right)
\left( P\pi_{\psi}(S_{j}^{(R)})^*)P - \pi_{\psi}(S_{j}^{(R)})^*)P \right) \Omega_{\psi})= 0, $$
we have
$$P\pi_{\psi}(S_{j}^{(R)})P = P\pi_{\psi}(S_{j}^{(R)})$$
and
$$ \sum_{j=1}^{d} \: v_{j} v_{j}^{*} =1.$$
Set $\tilde{v}_{j}  = J\Delta^{-1/2} v_{j}^{*} \Delta^{1/2} J$ and 
 $\tilde{v}_{j}^{*}  = J\Delta^{1/2} v_{j} \Delta^{-1/2} J$. 
The closure of $\tilde{v}_{j}$ and  $\tilde{v}_{j}^{*}$ are bounded operators satisfying
$$\sum_{j=1}^{d} \:  \tilde{v}_{j} \: \tilde{v}_{j}^{*} = 1$$
because
\begin{eqnarray*} 
&& \sum_{j=1}^{d}  ||  \tilde{v}_{j}Q \Omega_{\psi} ||^{2} =  
 \sum_{j=1}^{d} ||  Q \tilde{v}_{j}\Omega_{\psi} ||^{2} 
 \\ 
 &=&\sum_{j=1}^{d} ||  Q \pi_{\psi}(S_{j}^{(R)})^{*}\Omega_{\psi} ||^{2}
= \psi (\tau_{1} (Q^{*}Q)) = || Q \Omega_{\psi} ||^{2} 
\end{eqnarray*}
for $Q \in P \pi_{\psi}(O_{d}^{(R)})P$. Moreover,
\begin{equation}
 P\pi_{\psi}(S_{j}^{(L)})^{*}P = \tilde{v}_{j}^*.
\label{eqn:b13}
\end{equation}
(\ref{eqn:b13}) follows from the fact that  $\Omega_{\psi}$ is separating for 
the commutant of $E_{R}^{\prime} \frak N E_{R}^{\prime} = P \pi_{\psi}(O_{d}^{(R)})^{\dprime} P$ 
and the following equations:
\begin{equation}
P\pi_{\psi}(S_{j}^{(L)})^{*}P\Omega_{\psi} = P\pi_{\psi}(S_{j}^{(L)})^{*}\Omega_{\psi}  
= P\pi_{\psi}(S_{j}^{(R)})^{*}\Omega_{\psi} = \pi_{\psi}(S_{j}^{(R)})^{*}\Omega_{\psi} , 
\label{eqn:b14a}
\end{equation}
\begin{equation}
 \tilde{v}_{j}^*\Omega_{\psi}     = \pi_{\psi}(S_{j}^{(R)})^{*}\Omega_{\psi}  .
\label{eqn:b14b}
\end{equation}
\begin{lmm}
Let $\frak N$ be the von Neumann algebra on $\frak K$ generated by
$v_{j}$  as in Proposition \ref{pro:CuntzDual2}
and let $\frak N_{1}$ be the von Neumann algebra on $\frak K$ generated by
$\tilde{v}_{j}$. Then, 
\begin{equation}
{\frak N}^{\prime} ={\frak N}_{1} .
\label{eqn:b15}
\end{equation}
\label{lmm:MiniDual-a}
\end{lmm}
{\it Proof :}
${\frak N}^{\prime}$ is generated by $J{v}_{j}J$ and ${\frak N}_{1} \subset {\frak N}^{\prime}$ .
The modular operator $\Delta_{1}$ and the conjugation $J_{1}$ of ${\frak N}_{1} $ acting on 
$[{\frak N}_{1} \Omega_{\psi}]$ are the restriction of those for  ${\frak N}^{\prime}$ on $\frak K$.
Then $Jv_{j}J = \Delta^{-1/2} \tilde{v}_{j} \Delta^{-1/2}$ is in ${\frak N}_{1} $ .
{\it End of Proof.}
\begin{lmm}
\begin{equation}
[E_{R}\pi_{\psi}(O_{d}^{(R)})\Omega_{\psi}] = [E_{L}\pi_{\psi}(O_{d}^{(L)})\Omega_{\psi}] 
\label{eqn:b16}
\end{equation}
\label{lmm:MiniDual-b}
\end{lmm}
{\it Proof :}
By Lemma \ref{lmm:MiniDual-a},
the commutant of $\frak N = E_{R}\pi_{\psi}(O_{d}^{(R)})^{\dprime}E_{R}$ acting on 
$\frak K$ is $P \pi_{\psi}(O_{d}^{(L)})^{\dprime}P$. Obviously 
$P = E_{R} E_{R}^{'} \leq E_{R}E_{L} $. Then,
\begin{eqnarray}
&&[E_{R}\pi_{\psi}(O_{d}^{(R)})^{\dprime}\Omega_{\psi}] =
[P \pi_{\psi}(O_{d}^{(L)})^{\dprime}\Omega_{\psi}] 
\nonumber
\\
\subset &&
[E_{R}E_{L} \pi_{\psi}(O_{d}^{(L)})^{\dprime}\Omega_{\psi}]
= [E_{L}\pi_{\psi}(O_{d}^{(L)})^{\dprime}\Omega_{\psi}]
\label{eqn:b17}
\end{eqnarray}
The above inclusion tells us
$$[E_{R}\pi_{\psi}(O_{d}^{(R)})\Omega_{\psi}] \subset [E_{L}\pi_{\psi}(O_{d}^{(L)})\Omega_{\psi}] .$$
By the symmetry of $L$ and $R$ we have reverse inclusion. 
{\it End of Proof.}
\begin{lmm}
\begin{equation}
P = E_{R}E_{L}
\label{eqn:b18}
\end{equation}
\label{lmm:SupportProj}
\end{lmm}
{\it Proof:}
We show $P = E_{R}E_{R}^{\prime} \geq E_{R}E_{L}$.
 Due to (\ref{eqn:b10}), the Hilbert space $\frak H_{\psi}$ is generated by the vectors  
 $\pi_{\psi} (S_{I}^{(L)})  \pi_{\psi} (S_{J}^{(R)}) \pi_{\psi}(S_{J'}^{(R)})^{*} \Omega_{\psi}$.
 It suffices to show that the vector 
$\xi = E_{R}E_{L} \pi_{\psi} (S_{I}^{(L)})  \pi_{\psi} (S_{J}^{(R)}) \pi_{\psi}(S_{J'}^{(R)})^{*} \Omega_{\psi}$
 is in $\frak K$ (=the range of $P$).
 Due to the previous Lemma, $\eta = E_{L} \pi_{\psi} (S_{I}^{(L)})\Omega_{\psi} $ is in
 $\frak K$. Thus,
\begin{eqnarray}
&&E_{R}E_{L} \pi_{\psi} (S_{I}^{(L)})  \pi_{\psi} (S_{J}^{(R)}) \pi_{\psi}(S_{J'}^{(R)})^{*} \Omega_{\psi}
\nonumber\\
&=& 
E_{R} \pi_{\psi} (S_{J}^{(R)}) \pi_{\psi}(S_{J'}^{(R)})^{*} E_{L}\pi_{\psi} (S_{I}^{(L)})\Omega_{\psi}
\nonumber\\
&=& E_{R} \pi_{\psi} (S_{J}^{(R)}) \pi_{\psi}(S_{J'}^{(R)})^{*} \eta 
\nonumber\\
&=&E_{R} \pi_{\psi} (S_{J}^{(R)}) \pi_{\psi}(S_{J'}^{(R)})^{*} E_{R}\eta 
\in \frak K
\label{eqn:b19}
 \end{eqnarray}
{\it End of Proof.}
\bigskip
\noindent
\par
Now we return to proof of Proposition \ref{pro:CuntzDual2}.
First we look at the commutant of $E_{R}\pi_{\psi}(O_{d}^{(R)})^{\dprime}E_{R}$
on $E_{R}{\frak H}_{\psi}$. Obviously,
$E_{R}\pi_{\psi}(O_{d}^{(L)})^{\dprime}E_{R} \subset (E_{R}\pi_{\psi}(O_{d}^{(R)})E_{R})^{\prime}$
on $E_{R}{\frak H}_{\psi}$.
By Lemma \ref{lmm:MiniDual-a} and Lemma \ref{lmm:SupportProj},
$$E_{R}E_{L}\pi_{\psi}(O_{d}^{(L)})^{\dprime}E_{L}E_{R} = 
( E_{R}E_{L}\pi_{\psi}(O_{d}^{(R)})E_{L}E_{R})^{\prime} $$
Then, due to Lemma \ref{lmm:lifting2} 
$\pi_{\psi}(O_{d}^{(L)})^{\dprime} = \pi_{\psi}(O_{d}^{(R)})^{\prime}$ on $E_{R}{\frak H}_{\psi}$.
\\
Next we consider the inclusion 
$\pi_{\psi}(O_{d}^{(R)})^{\dprime} \subset \pi_{\psi}(O_{d}^{(L)})^{\prime}$ on ${\frak H}_{\psi}$.
As we already know that
$E_{R} \pi_{\psi}(O_{d}^{(R)})^{\dprime} E_{R} = E_{R}\pi_{\psi}(O_{d}^{(L)})^{\prime}E_{R}$
we apply Lemma \ref{lmm:lifting2} again and conclude that
 $\pi_{\psi}(O_{d}^{(R)})^{\dprime} = \pi_{\psi}(O_{d}^{(L)})^{\prime}$.
 {\it End of Proof of Proposition \ref{pro:CuntzDual2}.}
 \bigskip
 \bigskip
 \noindent
 \newline
{\bf Proof of Theorem \ref{th:duality}}
\newline
Now recall  Lemma \ref{lmm:extension-C}. We have two cases (i) and (ii). 
In the case (ii), our previous analysis shows that the pair of Cuntz algebras $\calB_{k}$
is in the von Neumann algebra $\pi_{\varphi}(\calA)^{\dprime}$ and the duality
follows from Proposition \ref{pro:CuntzDual2}.
\par
Hence we consider the case where the pure state $\psi$ of $\calB$ is invariant 
under $\gamma^{d}_{z}$. Let $U_{z}$ be the unitary implementing $\gamma^{d}_{z}$
and satisfying $U_{z}\Omega_{\psi}= \Omega_{\psi}$. We use the previous notation 
in our proof for Proposition \ref{pro:CuntzDual2}. By the duality for Cuntz algebras
(Proposition \ref{pro:CuntzDual2}), $E_{R} = E_{L}^{'}$ , $E_{L} = E_{R}^{'}$.
$E_{R}$ commutes with $U_{z}$ due to  $\gamma^{d}_{z}$ invariance
of $\psi$. As a result, the support projection of $\varphi$ for $\pi_{\varphi}({\frak A}_{R})^{\dprime}$
is the $E_{R}$ restricted to $\frak H_{\varphi}$. So we use the same notation $E_{R}$ (resp. $E_{L}$)
 for the support projection of $\varphi$ for $\pi_{\varphi}({\frak A}_{R})^{\dprime}$ 
 (resp. $\pi_{\varphi}({\frak A}_{L})^{\dprime}$ ).
\par
To show Haag duality we proceed as before. Taking into account of
$P = E_{R}E_{L}$ and 
$\pi_{\varphi}({\frak A}_{L})^{\dprime}  \subset  \pi_{\varphi}({\frak A}_{R})^{\prime} $,
it suffices to show
\begin{equation}
P \pi_{\varphi}({\frak A}_{L})^{\dprime} P = P \pi_{\varphi}({\frak A}_{R})^{\prime}P .
\label{eqn:b20}
 \end{equation}
On $\frak K_{0} = P\frak H_{\varphi}$ we apply Tomita-Takesaki theorem.
$P \pi_{\varphi}({\frak A}_{R})^{\prime}P$ is generated by $J_{0}v_{I}v_{K}^{*}J_{0}$.
where $I$ and $K$ are multi-indices satisfying $| I |=|K|$ 
and $J_{0}$ is the restriction of $J$ to $\frak K_{0}$.
By Haag duality for Cuntz algebras, $Jv_{I}J$ and $Jv_{K}^{*}J$ are approximated
in strong operator topology by elements $w_{\alpha}$ and  $x_{\alpha}$
of $P\pi_{\psi}(O_{d}^{(L)})P$.
Using Fourier decomposition(with help of $U_{z}$) we
may assume that 
\begin{equation}
U_{z} w_{\alpha}U_{z}^{*} =  z^{| I |} w_{\alpha} , \quad 
U_{z} x_{\alpha}U_{z}^{*} =  z^{| K |} x_{\alpha} .
\label{eqn:b21}
 \end{equation}
As a consequence, $Jv_{I}v_{K}^{*}J$ is approximated by
by elements of $P\pi_{\psi}(O_{d}^{(L)})P \cap \{U_{z} | z\in U(1)\}^{\prime}$. 
Thus  $J v_{I}v_{K}^{*}J$ is contained in 
$$P\pi_{\psi}(O_{d}^{(L)})^{\dprime}P \cap \{U_{z} | z\in U(1)\}^{\prime} = 
P\pi_{\psi}(\frak A_{L})^{\dprime}P $$ 
on $\frak K$.
By taking restriction to  $\frak K_{0}$, we see (\ref{eqn:b20}).
{\it End of Proof.}


\end{document}